# Co-substituted BiFeO$_3$: electronic, ferroelectric, and thermodynamic properties from first principles

Shivani Grover[a], Keith T. Butler[b], Umesh V Waghmare[c], Ricardo Grau-Crespo[a*]

[a] *Department of Chemistry, University of Reading, Whiteknights, Reading RG6 6DX, UK.*

[b] *SciML, Scientific Computing Division, Rutherford Appleton Laboratory, Harwell OX11 0QX, UK.*

[c] *Theoretical Sciences Unit, Jawaharlal Nehru Centre for Advanced Scientific Research, Bangalore 560064, India.*

*\* Corresponding author. Email:* r.grau-crespo@reading.ac.uk

**Abstract**

Bismuth ferrite, BiFeO$_3$, is a multiferroic solid that is attracting increasing attention as a potential photocatalytic material, because the ferroelectric polarisation enhances the separation of photogenerated carriers. With the motivation of finding routes to engineer the band gap and the band alignment, while conserving or enhancing the ferroelectric properties, we have investigated the thermodynamic, electronic and ferroelectric properties of BiCo$_x$Fe$_{1-x}$O$_3$ solid solutions, with $0 < x < 0.13$, using density functional theory. We show that the band gap can be reduced from 2.9 eV to 2.1 eV by cobalt substitution, while simultaneously increasing the spontaneous polarisation, which is associated with a notably larger Born effective charge of Co compared to Fe cations. We discuss the interaction between Co impurities, which is strongly attractive and would drive the aggregation of Co, as evidenced by Monte Carlo simulations. Phase separation into a Co-rich phase is therefore predicted to be thermodynamically preferred, and the homogeneous solid solution can only exist in metastable form, protected by slow cation diffusion kinetics. Finally, we discuss the band alignment of pure and Co-substituted BiFeO$_3$ with relevant redox potentials, in the context of its applicability in photocatalysis.

# 1. Introduction

Multiferroic materials, which simultaneously exhibit two or more ferroic properties (ferromagnetism or anti-ferromagnetism, ferroelectricity, and ferroelasticity), are promising for a range of functional applications [1]. Bismuth ferrite (BiFeO$_3$) is among the few attractive multiferroic materials with both ferroelectric and (anti)ferromagnetic behaviour at room temperature. It has a high ferroelectric-to-paraelectric transition point (Curie temperature $T_C$ = 1103 K) and an antiferromagnetic-to-paramagnetic transition point also well above ambient temperature (Néel temperature $T_N$ = 643 K) [2]. The crystal structure (**Figure 1**) of BiFeO$_3$ at ambient conditions is rhombohedral (space group R3c), with lattice parameter $a$= 5.64 Å and rhombohedral angle α= 59.4° [3-5]. The off-centre displacements of the Fe and O atoms with respect to the Bi sub-lattice result in a large spontaneous polarisation along the pseudo-cubic [111] direction, primarily due to Bi translation along this direction [6, 7]. The structure is also piezoelectric at all temperatures below $T_C$ [8]. Its magnetic structure below $T_N$ is, in a first approximation, G-type antiferromagnetic, which means that each Fe$^{3+}$ spin is surrounded by antiparallel spins on nearest-neighbour Fe sites, leading to zero net moment. But there is actually a weak net magnetic moment per unit cell, which results from spin canting due to the symmetry breaking induced by the ferroelectric polarisation [9]. These properties make BiFeO$_3$ a promising room-temperature single-phase multiferroic material, with potential applications in data storage, spin valves, spintronics and sensors [2].

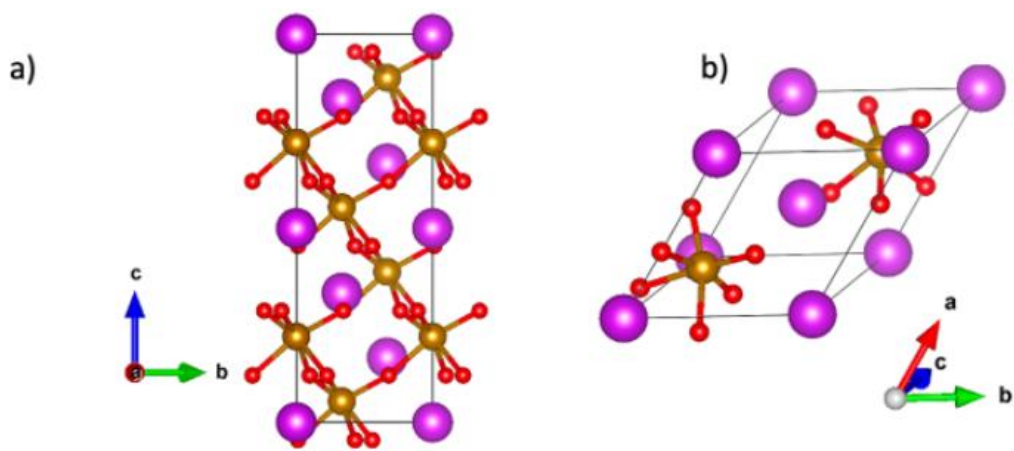

**Figure 1.** Crystal structure of BiFeO$_3$ in (a) hexagonal representation (conventional cell) and b) rhombohedral representation (primitive cell). Colour scheme: Bi = purple, Fe = brown, O = red.

Recently, there has also been increasing interest in using BiFeO$_3$ for photovoltaic [10-15] and photocatalytic [16-26] applications. This attention is motivated by the ferroelectric character of this oxide, because its internal electric field can be exploited to aid the separation of oppositely charged photogenerated carriers and/or to engineer the band alignment. Furthermore, BiFeO$_3$ has a relatively narrow band gap (most commonly reported values for thin films range between 2.6 and 2.8 eV [27]), compared to most other ferroelectric oxides, which makes it especially attractive for optical applications. Since this band gap is still somewhat too wide for optimal visible light absorption, there have been efforts to use chemical substitutions to make it narrower, while retaining or improving the ferroelectric properties of the oxide. Peng *et al.* [28] showed that Co/Fe substitution led to narrower band gaps (*e.g.* from 2.66 eV in pure BiFeO$_3$ to 2.53 eV in BiCo$_{0.1}$Fe$_{0.9}$O$_3$). Recent experimental work by Machado *et al.* [13, 14] has shown that, in addition to decreasing the band gap, Co/Fe substitutions can increase the spontaneous polarisation of BiFeO$_3$, which might be a bonus for photovoltaic and photocatalytic applications, by enhancing carrier separation.

We are particularly interested here in the applications of BiFeO$_3$ and BiCo$_x$Fe$_{1-x}$O$_3$ in photocatalysis. Despite having an adequate band gap, there is evidence that pure BiFeO$_3$ does not have the right band alignment for the full water splitting reaction and therefore is inactive as a single-semiconductor photocatalyst for that reaction [25], but it can still be used for water photo-oxidation biased by sacrificial agents (which are then reduced instead of the protons) [29]. BiFeO$_3$ also shows promising behaviour when used as part of heterojunctions for water splitting photocatalysis. For example, CdS/BiFeO$_3$ heterojunctions forming a Z-scheme have been found to be active as particulate photocatalysts for water splitting without using any sacrificial agents [25]. Furthermore, BiFeO$_3$-based materials can be used as photoanodes in tandem photoelectrochemical (PEC) cells, where the photoanode or photocathode only need to have suitable band alignment for one half reaction. Khoomortezaei *et al.* have demonstrated that photoanodes made of WO$_3$/BiVO$_4$/BiFeO$_3$ [22] or WO$_3$/BiFeO$_3$ [24] heterojunctions are efficient in photo-electrochemical water splitting. BiFeO$_3$ can also be useful as photocatalyst (or part of them) for other reactions, such as the degradation of organic pollutants (see Ref. [19] for a review). For example, studies involving both pure [17] and doped [23] BiFeO$_3$, as well as BiFeO$_3$–containing heterojunctions (*e.g.* with C$_3$N$_4$ [20]) have shown that BiFeO$_3$ is an active photocatalyst for degrading rhodamine B, an organic dye that is widely used as a colorant in the textile industry, and can be toxic to humans and animals if not removed from wastewaters. Mushtaq *et al.* have shown that the photocatalytic activity of BiFeO$_3$ in the degradation of rhodamine B can be enhanced under mechanical vibrations thanks to piezo-photocatalytic effects [21]. In these applications, the ferroelectric character of $BiFeO_3$ is beneficial to the photocatalytic process by aiding the separation of photogenerated carriers. Recent work by Huang *et al.* [26] has shown that the effect can be enhanced by electrically poling (via the application of an external electric field) the $BiFeO_3$ nanoparticles to align the ferroelectric domains, which accelerated the photocatalytic process by a factor of two compared with unpoled $BiFeO_3$. There is also evidence that the orientation of the ferroelectric polarisation can be used to engineer photocatalytic response [18, 30]. The interplay between ferroelectric and photocatalytic properties in this material clearly deserves further research attention at the theoretical level to improve our fundamental understanding of these phenomena, which will help rationalising the design of better $BiFeO_3$-based photocatalysts.

In this work, we theoretically investigate the incorporation of cobalt in $BiFeO_3$, as a route to engineer its band gap, band alignment, and ferroelectric polarisation. We discuss the thermodynamic aspects of the Co/Fe substitutions, in particular the stability of the solid solution with respect to phase separation. We examine the impact of cobalt substitution on the magnitude of the $BiFeO_3$ spontaneous polarisation, which can improve charge separation and can potentially be used to engineer the band alignment for photocatalysis. We calculate the depolarisation field in a $BiFeO_3$ thin film with the surface normal parallel to the polarisation direction, and discuss the interplay between polarisation and band alignment. In the light of our simulation results, we also discuss conflicting experimental results in the literature about the band gap, band alignment, and polarisation of these materials.

## 2. Computational details

Our calculations are based on the density functional theory (DFT) as implemented in the Vienna Ab Initio Simulation Package (VASP) [31, 32]. For relaxations, substitution thermodynamics, and for the electronic structure calculations, we used a screened hybrid functional, based on the functional by Heyd, Scuseria and Ernzerhof (HSE06) [33] which admixes the exact non-local exchange from the Hartree-Fock theory, screened at long-range with a screening parameter 0.2 Å$^{-1}$, into the local Perdew-Burke-Ernzerhof functional [34] of the generalized gradient approximation (GGA). In our case, 20% Hartree-Fock exchange (instead of the 25% proposed in the original HSE06 functional) was chosen, following Shimada *et al.* [4], who showed that these settings led to good agreement with experiment in terms of both structural parameters and band gap. In what follows we refer to the screened hybrid functional used here simply as HSE.

For the calculation of Born effective charges and polarisations, which is computationally demanding, we used the less expensive GGA+U approach, where Hubbard effective parameters ($U_{\text{eff}}$) of 4 eV and 3.3 eV are applied to the 3d orbitals of Fe and Co, respectively; these values were originally fitted to reproduce the oxidation energies of the corresponding binary transition metal oxides [35], and are found to transfer well to describe the properties of more complex oxides of these metallic elements [36-38].

The projector augmented wave (PAW) method [39, 40] was used in all calculations to describe the interactions between the valence electrons and frozen cores, by explicitly treating 15 valence electrons for Bi, 8 for Fe, 9 for Co, and 6 for oxygen. We used an energy cutoff of 520 eV to truncate the plane wave expansion of the Kohn-Sham wavefunctions, which is 30% above the default cutoff for the employed PAW potentials, to minimise Pulay errors. Brillouin zone (BZ) integrations were performed by sampling the reciprocal space using a $\Gamma$−centred mesh of $4 \times 4 \times 4$ k-points with reference to the rhombohedral unit cell, and commensurate grids for supercells. The only exception were the structural relaxations, using the HSE functional, of the different Co-substitution configurations in the $2 \times 2 \times 2$ supercell, which were performed using Γ-only calculations. In all cases, the cell parameters were allowed to relax, and the ions were moved towards equilibrium until the Hellmann-Feynman forces were less than 0.01 eV/Å.

The DFT calculations were all spin-polarised, and the G-type antiferromagnetic ordering was assumed, as well as collinear spin arrangements, *i.e.* the small effect of spin canting in $BiFeO_3$ was ignored. Both Fe(III) and Co(III) cations were initialised in high-spin (HS) configurations, which the calculations conserved after convergence. Iron in $BiFeO_3$ is well known to be in HS Fe(III) configuration, unless high pressures are applied, in which case a transition to low-spin (LS) takes place [2, 41]. The spin state of Co (III) (a $d^6$ cation) in $BiCo_xFe_{1-x}O_3$ is more disputed in the literature. In the pure-Co end-member, $BiCoO_3$, Co(III) is in HS state at ambient pressure and low temperatures [42]. But Ray *et al.* concluded, from magnetisation measurements, that dilute Co in $BiFeO_3$ at low temperature (below 150 K) and low external magnetic field, was LS [43]. On the other hand, from a combination of magnetic measurements and DFT calculations, Fan *et al.* concluded that at ambient temperature the LS state of Co(III) in this system is unfavoured, and their GGA+U calculations indicated that HS

Co(III) was more stable than LS Co(III) in BiFeO$_3$ [44]. Our own test calculations using the HSE functional in the supercell with one Co substitution showed that HS Co(III) is more stable than LS Co(III) in BiFeO$_3$, by ~0.2 eV. Therefore, we have used HS Co(III) in all our calculations.

The symmetrically distinct Co substitution configurations were found using the SOD (Site Occupancy Disorder) program [45, 46]. Two configurations are considered equivalent if they are related by a symmetry operator, and the group of symmetry operators of the supercell consists of the original symmetries of the unit cell and their combinations with supercell translations. The DFT energies of the symmetrically independent configurations were used to fit an effective Hamiltonian model, which we used to perform Monte Carlo simulations in a large (10 × 10 × 10) supercell with the same composition, BiCo$_{0.125}$Fe$_{0.875}$O$_3$, using the GULP code [47, 48]. At each step of Monte Carlo simulation, a configuration is created by randomly swapping a pair of atoms, which corresponds to an energy difference $\Delta E$. The new configuration is accepted, following the Metropolis algorithm [49], with probability $p = \min(1, \exp(-\Delta E/k_\mathrm{B} T))$, where $k_\mathrm{B}$ is Boltzmann's constant and $T$ is the temperature. Forty million steps were used to achieve equilibrium, and the simulations were performed at $T$=500 K, 1000 K and 2000 K.

For the alignment of the band edges of BiFeO$_3$ with respect to vacuum, we built periodic slab models with different terminations, separated by a vacuum gap of 15 Å in the periodic supercell. We use the MacroDensity code [50] for calculating the planar averages of the electrostatic potential in planes parallel to the slab surface, to determine the potential difference between the bulk (average in the middle of the slab) and the vacuum level (in the middle of the vacuum gap). For the alignment of the band edges of BiCo$_{0.06}$Fe$_{0.94}$O$_3$ we did not create slabs with Co/Fe substitutions, but simply aligned the core levels of bulk Fe atoms far from Co substitutions.

## 3. Results and discussion

### 3.1 Structural and electronic properties

Calculated structural and electronic parameters of pure BiFeO$_3$ are shown in **Table 1** in comparison to experimental values. As noted before in Ref [4], the screened hybrid functional HSE leads to good agreement with experiment in terms of both crystal structure and band gaps.

The GGA+U predictions are also reasonable, albeit with a band gap below the most accepted experimental range of 2.6-2.8 eV measured at ambient conditions [27]. Our HSE band gap is slightly above that range. However, it should be noted that there is still a lot of uncertainty about the band gap of BiFeO$_3$, with values as low as ~2 eV or as high as ~3 eV reported in the experimental literature for this compound, depending on synthesis conditions, resulting morphology or particle size, and measurement method [27, 51]. Smaller nanoparticles tend to have narrower band gaps [52]. The band gap of BiFeO$_3$ also decreases substantially with temperature, *e.g.* from about 2.5 eV at ambient temperature to about 1.5 eV at 550 °C in Ref. [53]. The value of 2.9 eV obtained here for bulk BiFeO$_3$ from HSE calculations seems therefore to be a more reasonable zero-Kelvin prediction than the GGA+U value of 2.3 eV.

**Table 1.** Calculated lattice parameters, $a$ and $\alpha$, and band gap $E_g$ for pure BiFeO$_3$ (R3c phase). Experimental values (obtained at ambient temperature) for the cell parameters [5] and for the band gap [27] are listed for comparison.

| Parameter | Expt. | GGA+U | HSE |
|---|---|---|---|
| $a$ / Å | 5.64 | 5.69 | 5.68 |
| $\alpha$ / deg | 59.42 | 59.0 | 58.8 |
| $E_g$ / eV | 2.6-2.8 | 2.3 | 2.9 |

The band structure (**Figure 2a**) and density of states (DOS) (**Figure 2b**) show that the contribution of O 2p orbitals dominate the valence band (VB) edge and the contribution of Fe 3d orbitals dominate the conduction band (CB) edge. The bottom of the CB is located at the Z point of the Brillouin zone. The VB has two maxima at roughly the same energy, one between Γ and F, and the other at the Z point (the difference between the two is less than 10 meV). Both spin channels exhibit the same total density of states due to the antiferromagnetic arrangements of the magnetic moments (Fe 3d contributions with opposite spins at the same energy level come from different atoms). There is a large exchange splitting of ~9 eV between the occupied and empty 3d orbitals of a given Fe atom, therefore the occupied Fe 3d orbitals are not shown in the DOS plot.

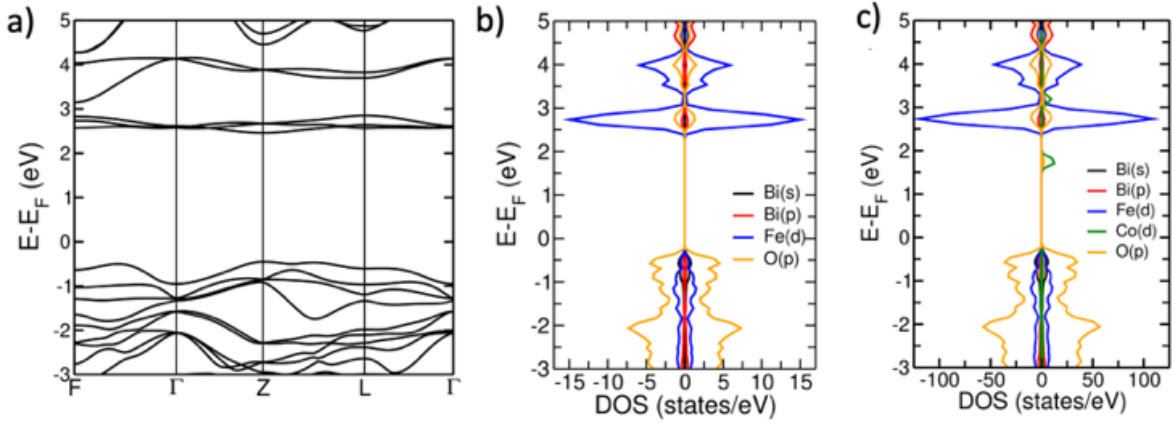

**Figure 2.** a) Band structure and (b) projected density of states of pure BiFeO$_3$; and c) projected density of states of BiCo$_{0.06}$Fe$_{0.94}$O$_3$.

We then considered the substitution of one or two Fe atoms in the $2 \times 2 \times 2$ supercell by Co, in such a way that the stoichiometry of the solid solution is BiCo$_{0.062}$Fe$_{0.938}$O$_3$ or BiCo$_{0.125}$Fe$_{0.785}$O$_3$. These compositions are within the experimentally observed range of stability ($0 < x < 0.2$) of the rhombohedral structure for the BiCo$_x$Fe$_{1-x}$O$_3$ solid solution (upon increasing $x$, the solid solution transitions to the tetragonal structure of pure BiCoO$_3$, via an intermediate monoclinic structure [54]). The DOS plot for BiCo$_{0.062}$Fe$_{0.938}$O$_3$ (**Figure 2c**) shows the presence of an empty impurity level of Co 3$d$ character that reduces the band gap from 2.9 to 2.1 eV.

The supercell with composition BiCo$_{0.125}$Fe$_{0.875}$O$_3$ (two Co substitutions in the supercell) has only five symmetry inequivalent configurations. The DOS plots for those configurations (**Figure 3**) are very similar to the DOS plots for lower Co concentration, but with larger peaks in the Co 3d gap state, whose energy position do not change appreciably with the relative position of the Co atoms in the structure. In these calculations, the magnetic moments of the Co ions were given the same orientation of the magnetic moments of the Fe ions they replace, thus keeping the G-type antiferromagnetic pattern; this accounts for the different relative orientations of the Co magnetic moments over the configurations (*i.e.* they are parallel in the configurations shown in **Figure 3 a, b, c**, but antiparallel in **d** and **e**). In the next section we use the relative energies of these configurations to create a model to investigate the equilibrium distribution of Co substitutions.

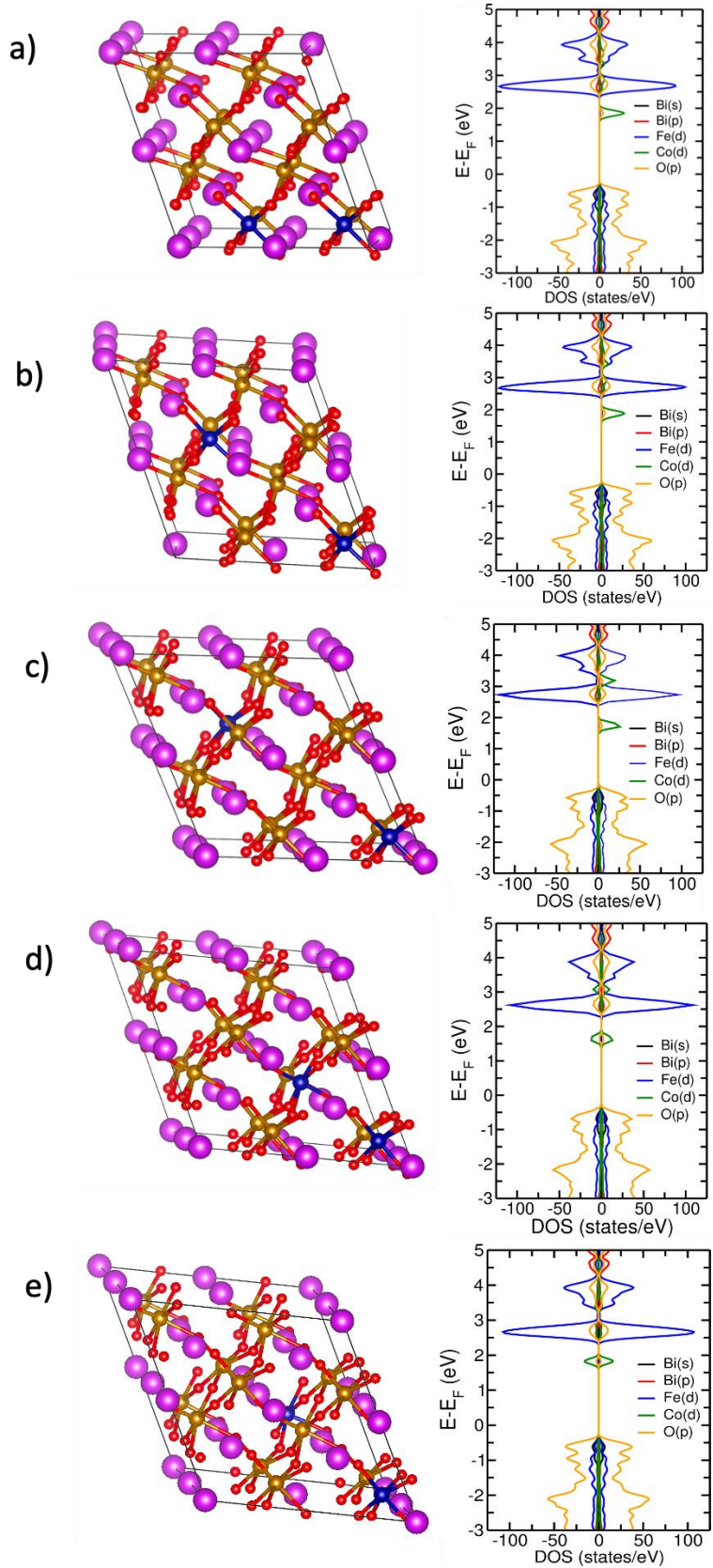

**Figure 3.** Co positions and projected density of states for the five symmetrically distinct configurations of BiCo$_{0.06}$Fe$_{0.94}$O$_3$ studied in this work. Colour scheme: Bi = purple, Fe = brown, Co = blue, O = red.

## 3.2 Thermodynamics of cation distribution

For the thermodynamic analysis of the Co distribution, the DFT total energies of the five distinct configurations of double substitutions were mapped into an Ising-like Hamiltonian:

$$E = E_0 + \sum_{<i,j>} J_{ij} S_i S_j \qquad (1)$$

where the "spin" variable $S_i = 1$ if site $i$ is occupied by Co, and $S_i = 0$ if it is occupied by Fe, and the $J_{ij}$ values characterise the strength of Co-Co interactions. Note that the spin variables and the Hamiltonian itself are unrelated to the magnetism of the system but simply describe the interactions between Co impurities. Four *J* constants, corresponding to four different Co-Co distances, plus the $E_0$ values, are then obtained from the five DFT energies, by solving the system of five linear equations with five variables.

The interaction parameters *J* as a function of Co-Co distance (*d*) are shown in **Figure 4a**. Clearly, the interaction between impurities is more attractive when the distance between them is shorter. To study the effect of these interactions on the equilibrium cation distribution at specific temperatures, we performed Monte Carlo simulations in a larger ($10 \times 10 \times 10$) supercell with the same composition, $BiCo_{0.125}Fe_{0.875}O_3$. At *T*= 500 K, Co impurities aggregate as one pure-Co spherical cluster per simulation cell (**Figure 4b**). The formation of such small separate Co clusters is, of course, an artifact from the simulation cell size; using a larger simulation cell would lead to larger and more separated Co clusters. The result from the Monte Carlo simulation simply indicates that there is a thermodynamic preference for Co aggregation, which would lead to complete phase separation at that temperature. Our simulations at $T = 1000$ K and 2000 K still led to Co impurity aggregation forming single clusters, albeit with a less compact shape and not so well-defined borders. Only at unrealistically high temperatures does the equilibrium cation distribution becomes more homogeneous.

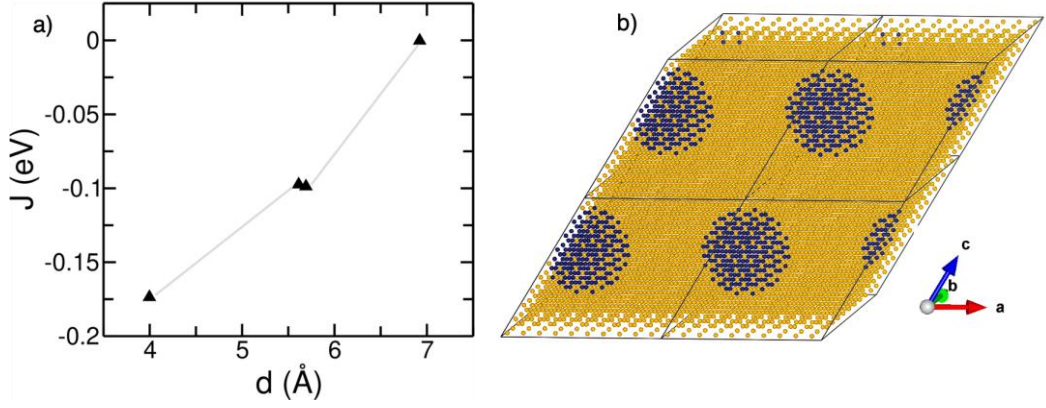

**Figure 4**. (a) Interaction parameter ($J$) as a function of Co-Co distance ($d$) for 12.5% Co concentration. The grey line is a guide to the eye. (b) Equilibrated distribution of impurities after Monte Carlo simulations at $T=$ 500 K. Only Co atoms (in blue) and Fe atoms (in brown) are shown.

While these simulations show that there is a strong thermodynamic drive towards phase separation, they implicitly force the BiCoO$_3$ clusters to remain in the rhombohedral structure of BiFeO$_3$. In reality, it would be thermodynamically favourable for BiCoO$_3$ to separate forming its preferred crystal structure, which is tetragonal [54]. To quantify the thermodynamic stability of the BiCo$_x$Fe$_{1-x}$O$_3$ solid solution against phase separation to rhombohedral BiFeO$_3$ and tetragonal BiCoO$_3$ phases, we have calculated the enthalpy of mixing per formula unit of BiFe$_{1-x}$Co$_x$O$_3$ with respect to the pure compounds using the equation:

$$\Delta H_{\text{mix}} = E[\text{BiFe}_{1-x}\text{Co}_x\text{O}_3] - (1-x)E[\text{BiFeO}_3] - xE[\text{BiCoO}_3], \qquad (2)$$

where the $E$ values are the DFT energies per formula unit for the corresponding compositions, at their ground-state structures. For $x = 0.0625$, we obtain $\Delta H_{\text{mix}} = 0.027$ eV per formula unit. As is common in the description of very dilute solid solutions (*e.g.* [55, 56]), we can write that for small values of $x$:

$$\Delta H_{\text{mix}}(x) = Wx, \qquad (3)$$

where $W = 0.43$ eV is the solution energy. From this value we can estimate the thermodynamic solubility limit $x_s$ of Co in BiFeO$_3$, as the minimum of the free energy of mixing (including $\Delta H_{\text{mix}}(x)$ and the configurational entropy contribution) at a given temperature, which is:

$$x_s \approx e^{-\frac{W}{k_B T}}. \qquad (4)$$

For example, at $T = 600$ K, $x_s = 0.00026 = 260$ ppm. Therefore, Co substitution in BiFeO$_3$ can be expected to be thermodynamically stable against phase separation only at trace amounts.

However, it is important to realise that even when phase separation is thermodynamically preferred, homogeneous solid solutions can still exist, protected by the very slow kinetics of cation diffusion. Cation exchange barriers are well known to be very high, typically ~2 eV or above for ionic solid solutions. For example, values of 193 kJ/mol, 200 kJ/mol, and 230 kJ/mol have been estimated from experimental measurements of cation exchange kinetics in (Fe$_{0.5}$Mn$_{0.5}$)$_2$SiO$_4$ olivines [57], (Fe$_3$O$_4$)$_x$(MgAl$_2$O$_4$)$_{1-x}$ magnetite-spinel solid solutions [58], and disordered MgAl$_2$O$_4$ spinels [59], respectively. These high activation barriers mean that cation exchange typically only starts, at any measurable rate, if samples are heated above ~700 K, whereas full equilibrium (involving either ex-solution or ordering) is only reached at much higher temperatures. For Co-substituted BiFeO$_3$, we have not calculated cation exchange activation barriers (such calculations are not trivial, because the mechanism for cation exchange may involve the collective movements of many atoms and/or be mediated by vacancies or other defects), but it can be safely expected that they would be similarly high. Therefore, while there is clearly a thermodynamic driving force for ex-solution in Co-substituted BiFeO$_3$, such separation process would only be observable if samples are treated at high temperatures. When a homogeneous solid solution is prepared it should remain stable (or more precisely, metastable) if kept at ambient or only moderately high temperatures. An example of such metastable oxide solid solution used as a functional material is Ce$_{1-x}$Zr$_x$O$_2$ [60], which is thermodynamically unstable with respect to phase separation into Ce-rich and Zr-rich oxides at most compositions, but still can be synthesised as a homogeneous solid solution that is widely used in catalysis at moderate temperatures [61, 62].

### 3.3 Ferroelectric properties

To study the effect of cobalt substitution on ferroelectric properties of BiFeO$_3$, we calculate the spontaneous polarisation $P$ from first principles. The Born effective charge (BEC) tensor for atom $j$, ($Z^*_{j,\alpha\beta}$) is defined as the derivative of the polarisation $P$ with respect to the atom coordinates (at zero electric field):

$$Z^*_{j,\alpha\beta} = \left.\frac{\partial P_\alpha}{\partial u_{j,\beta}}\right|_{\epsilon=0} \quad (5)$$

where $\alpha$ and $\beta$ are Cartesian indices, and can be obtained in VASP using density functional perturbation theory [63]. Then, to estimate spontaneous polarisation we compute the Cartesian components of the polarisation $P_\alpha$ as:

$$P_\alpha = \frac{e}{V} \sum_{j\beta} \bar{Z}^*_{j\alpha\beta} \Delta u_{j\beta} \quad (6)$$

where

$$\bar{Z}^*_{j\alpha\beta} = \int_0^1 Z^*_{j\alpha\beta}(\xi) d\xi \quad (7)$$

is the average of the BEC tensor over the values of the distortion parameter $\xi$ that connects the ferroelectric R3c phase ($\xi = 1$) and the reference, centrosymmetric R-3c phase ($\xi = 0$); $\Delta u_{j\beta}$ is the displacement of ion $j$ in the Cartesian direction $\beta$. The polarisation calculated at different distortion points is shown in **Figure 5**.

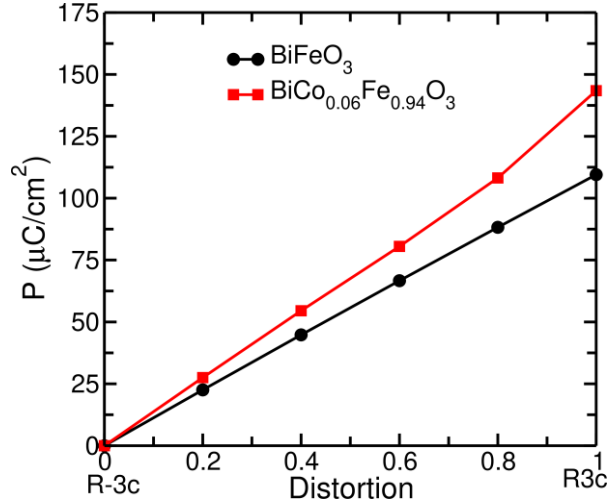

**Figure 5**. Change in the polarisation [111] component along a path from the centrosymmetric R-3c structure to polar R3c phase for $BiFeO_3$ and $BiCo_{0.06}Fe_{0.94}O_3$.

For pure $BiFeO_3$, we obtained 109 $\mu C/cm^2$, and it increases to 143 $\mu C/cm^2$ for $BiCo_{0.06}Fe_{0.94}O_3$ (there is only one Co atom per supercell at this composition, and there is only one symmetrically distinct Fe site, so the result does not depend on the substitution site). Let us first discuss the value for the pure compound. It is well known that experimental measurements of polarisation in this material have led to a very wide spread of results (*e.g.* Table I in Ref. [7] lists experimental values from different sources ranging from 2.2 to 158 $\mu C/cm^2$). There are several causes for this spread of polarisation values measured in experiment. Polarisation measurements are often affected by leakage current problems, which might account for some of the very small values reported for $BiFeO_3$. Poor sample quality, or the presence of structural modifications, can also affect the result. Furthermore, the spread of polarization values may also be related to the multi-valued nature of that property, which is well explained by the modern theory of polarization [64]: polarisation in solids is in fact a "lattice" of values, rather than a single vector. Very different values can then be obtained if the structural paths of ferroelectric switching measurements in samples are substantially different. In Ref. [7] it was concluded that the most "natural" value of the polarisation of pure $BiFeO_3$ along the <111> direction should be 90-100 $\mu C/cm^2$, which was consistent with the most recent thin film measurements. A value of 100 $\mu C/cm^2$ is also quoted in the review by Catalan and Scott, as a sensible value [2]. Our theoretical result of 109 $\mu C/cm^2$ for pure $BiFeO_3$, is close to these best estimations.

To compare the predicted effect of Co substitution on the $BiFeO_3$ polarisation, we turn to the experimental results presented by Machado *et al.* in Ref. [14]. In that work, pure $BiFeO_3$ was measured to have a polarisation of only 26.1 $\mu C/cm^2$, which is low compared to currently accepted values. Therefore, we cannot directly compare our calculated absolute values with their experiments. However, the qualitative effect of Co substitution can be compared: they observed an increase in the polarisation to 60 $\mu C/cm^2$ for $Co_{0.1}Bi_{0.9}FeO_3$, in comparison with the pure compound. In our calculations, the behaviour is the same: the incorporation of Co to $BiFeO_3$ significantly enhances its ferroelectric polarisation. The greater BEC of Co compared to Fe (**Table 2**) is consistent with the enhanced polarisation in the substituted system. Interestingly, in Ref. [13], a GGA+U calculation with a higher value of the Hubbard parameter for Co 3d orbitals, $U_{eff}$=6 eV, compared to the value used in this work, $U_{eff}$=3.3 eV, did not lead to much difference in polarisation between pure and Co-substituted $BiFeO_3$. We have not been able to reproduce their results exactly (there are other differences in methodology), but we do observe a much smaller enhancement (~50% of ours) of the polarisation upon Co substitution when we use their $U_{eff}$ value. The effect of $U_{eff}$ and other calculation settings on the estimated polarisation deserves further study.

**Table 2.** Born effective charges (BECs) (average of diagonal elements for each atom), and polarisation *P*, for $BiFeO_3$ and $BiCo_{0.06}Fe_{0.94}O_3$.

|  | BEC (*e*) | | | $P(\mu C/cm^2)$ |
|---|---|---|---|---|
|  | Bi | Fe | Co |  |
| $BiFeO_3$ | 4.64 | 3.97 |  | 109 |
| $BiCo_{0.06}Fe_{0.94}O_3$ | 4.97 | 4.02 | 5.24 | 143 |

## 3.4 Band alignment and applications in photocatalysis

To explore the activity of these compounds in photocatalytic applications, it is important to estimate the positions of the VB maximum and the CB minimum in an absolute scale, for example, relative to the vacuum level. This is necessary to obtain the relative energies of these band edges with respect to half-reaction potentials for the redox reactions of interest, *e.g.* water splitting. However, energy levels obtained from the bulk calculation are given relative to the average electron potential in the solid. To align the electronic structure with reference to the vacuum level, we determine the potential difference $\Delta V$ between the vacuum potential and the pseudo-bulk average using an auxiliary slab calculation. In the first instance, to ignore the effect of ferroelectric polarisation, we have used a symmetric and stoichiometric slab terminated by the non-polar (110) surface, for which the vacuum level is the same at both sides of the slab (**Figure 6**).

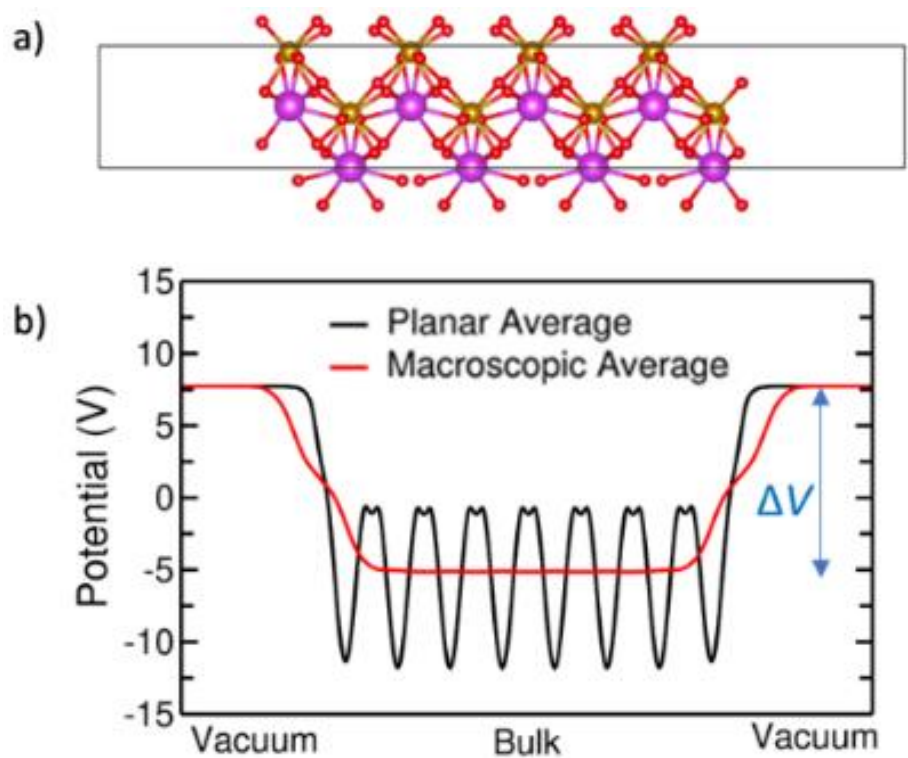

**Figure 6**: (a) BiFeO$_3$ slab along with (110) surface termination and (b) planar average electrostatic potential for that slab.

The calculated positions of the band edges of BiFeO$_3$ and BiCo$_{0.06}$Fe$_{0.94}$O$_3$ with respect to vacuum potential are shown in **Figure 7**. We can compare these with the half-reaction potentials for water splitting, corresponding to the hydrogen evolution reaction (HER) and the oxygen evolution reaction (OER):

$$2H^+_{(aq)} + 2e^- \leftrightarrow H_{2(g)} \quad \text{(HER)} \tag{8}$$

$$H_2O \leftrightarrow 2H^+_{(aq)} + \frac{1}{2}O_{2(g)} + 2e^- \quad \text{(OER)} \tag{9}$$

In the vacuum scale and at pH $= 0$, the HER level is located at -4.44 eV, and the OER level is located at -5.67 eV. At temperature $T$ and pH $> 0$, these energy levels are shifted up by $k_B T \times \text{pH} \times \ln 10$. For pH $= 7$ and room temperature, the HER level is then located at -4.03 eV and OER level is located at -5.25 eV, respectively. For a water splitting photocatalyst made of a single semiconductor, the positions of the CB minimum and the VB maximum should straddle those half-reaction potentials. The band edges of BiFeO$_3$ and BiCo$_{0.06}$Fe$_{0.94}$O$_3$ and the water redox potentials are shown in **Figure 7.** The calculated band edges are not well aligned with the half-reaction potentials for water splitting, because the CB edges are too negative with respect to the HER level, suggesting that these materials are not suitable as single-semiconductor photocatalysts for the full water splitting reaction. However, according to this result, BiFeO$_3$ can be used as a photoanode in a PEC cell, for the water oxidation reaction.

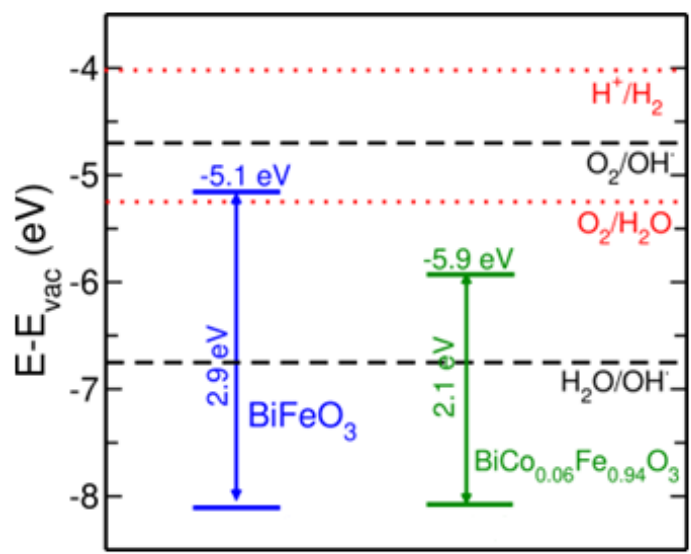

**Figure 7**. CB and VB edges calculated with the HSE functional for BiFeO$_3$ and BiCo$_{0.06}$Fe$_{0.94}$O$_3$. The half-reaction potentials for water splitting are represented by dotted red lines, and those for the generation of ȮH radicals are represented by dashed black lines.

When comparing our band alignment with experiment, we must recall that to build band alignment diagrams, experimentalists measure or estimate at least two out of four properties: the ionization potential (IP), the electron affinity (EA), the band gap ($E_g$) and/or the electronegativity ($\chi$). The relation between these quantities and the band edges $E_{CB}$ and $E_{VB}$ is given by the equations:

$$E_{CB} = -EA = -\chi + \tfrac{1}{2} E_g, \tag{10}$$

$$E_{VB} = -IP = -\chi - \tfrac{1}{2} E_g. \tag{11}$$

Usually, the optical band gap is measured, and one of the other three properties is also obtained, to complete the band diagram. Given that there are wide differences in measured band gaps, and that there are various ways to measure or estimate the other properties, strong discrepancies in the band alignments reported for BiFeO$_3$ in the experimental literature are not surprising. For example, Ji *et al.* [18] calculated the IP of BiFeO$_3$ from the ultraviolet photoelectron spectrum (by subtracting the width of the spectrum from the exciting photon energy). Combined with a band gap of 2.74 eV, that measurement led to a CB edge at -3.56 eV and a VB edge at -6.30 eV in the vacuum scale, which *do* straddle the water splitting redox potentials. A similar band alignment for BiFeO$_3$ band edges with vacuum had been proposed before by Wu and Wang [65]. These authors measured an optical band gap of 2.73 eV, and estimated the EA at 3.33 eV following Clark and Robertson [66], who had given that value from a simple comparison with SrBi$_2$Ta$_2$O$_9$. That leads to a CB edge at -3.33 eV and a VB edge at -6.06 eV in the vacuum scale, also straddling the water splitting redox potentials. However, the EA of BiFeO$_3$ seems to be significantly underestimated in those studies. Measurements by Moniz *et al.* [29] using electrical impedance spectroscopy and a Mott-Schottky plot [67] led to an EA of 4.62 eV. That value means that the band edges of BiFeO$_3$ *do not* straddle the water splitting redox potentials, the CB edge being too negative. This result also agrees with the simpler estimation of band positions made by Li *et al.* [16], using the electronegativity $\chi$ of the oxide calculated as the geometric mean of the electronegativities of the constituent atoms, and a measured band gap of 2.19 eV, from where they arrive at an EA of 4.94 eV. The same method has been applied by other authors, using slightly different band gaps [17, 25]. Kolivand and Sharifnia [25] confirmed the value determined by this method (EA = 4.84 eV) with their own Mott-Schottky plot analysis. Interestingly, if we use our HSE band gap, the average electronegativity method leads to almost the same band positions that we obtain from the HSE calculation in the auxiliary slab.

Table 3. Estimations of the electron affinity (EA) of $BiFeO_3$ reported in the literature, in comparison with the values obtained in this work. In some of the original papers, values are reported with respect to the normal hydrogen electrode (NHE), and have been converted here to the vacuum scale for easy comparison.

| Source | EA (eV) | Method |
|---|---|---|
| Clark & Robertson (2007) [66] <br> Wu & Wang (2010) [65] | 3.33 | Comparison with $SrBi_2Ta_2O_9$ |
| Li *et al.* (2009) [16] | 4.94 | Band gap measured and electronegativity estimated from elements |
| Ji *et al.* (2013) [18] | 3.56 | Band gap measured and ionization potential from ultraviolet photoelectron spectrum |
| He *et al.* (2013) [17] | 5.01 | Band gap measured and electronegativity estimated from elements |
| Moniz *et al.* (2014) [29] | 4.62 | Electrical impedance spectroscopy and Mott-Schottky plot |
| Kolivand & Sharifnia (2020) [25] | 4.84 | Band gap measured and electronegativity estimated from elements; Mott-Schottky plot |
| This work | 5.13 | DFT (HSE) calculations in bulk and auxiliary slab |

**Table 3** summarises EA estimations from the literature in comparison with our theoretical results. Our results agree with the higher EA estimations in the literature from Mott-Schottky measurements (albeit with some overestimation). Thus, our findings give support to the conclusion that unmodified $BiFeO_3$ cannot be used as a single-semiconductor photocatalyst for water splitting, because the EA is too high (the CB is too negative) to drive the hydrogen evolution half-reaction. Cobalt substitution does not help on this issue, because the low-lying empty Co 3d levels increase the EA and lower the CB further, as seen in **figure 7**. However, as mentioned earlier, $BiFeO_3$-based systems can be useful as photoanodes in PEC cells, because of the highly oxidising VB holes; effective coupling with a photocathode for the HER could make an efficient tandem system. $BiFeO_3$ can also be used for water splitting in heterojunctions with other semiconductors like CdS, as demonstrated by Kolivand and Sharifnia [25]. They showed that pure $BiFeO_3$ was not photocatalytically active for the full water splitting reaction, which is consistent with our theoretical band alignment. However, when $BiFeO_3$ is in a heterojunction with CdS, which has a lower EA (they estimate ~4 eV), a direct Z-scheme [68] band alignment forms, which allows unassisted water splitting without using any sacrificial agents. In principle, for such heterojunctions, Co substitution in $BiFeO_3$ might be advantageous, because the band gap narrowing with respect to the pure compound would allow more efficient visible light absorption at the $BiCo_xFe_{1-x}O_3$ end of the heterojunction, while the more negative CB in the solid solution, compared to the pure oxide, would reduce the losses associated to the interfacial electron-hole recombination. However, our results suggest that the CB in the Co-substituted system might be too low to drive the water oxidation reaction, although our calculation seems to somewhat overestimate the EA values, so it is difficult to establish a firm conclusion. Furthermore, it is unclear to what extent Co impurities could act as a trap state for photogenerated electrons, which would be detrimental for photocatalytic activity.

Our analysis above has focused on water splitting, but $BiFeO_3$-based systems could also be used as photocatalysts for other reactions, potentially even without help from heterojunctions. For example, the production of hydroxyl radicals from water for oxidation of organic pollutants might be a more suitable reaction [69]. These ȮH radicals can be useful for the degradation of effluents from textile and pharmaceutical industries. The relevant redox pairs and the energy levels of the half reactions are also shown in **figure 7.** Our calculated band edges for $BiFeO_3$ do not quite straddle the half-reaction potentials, but because our EA value is likely to be slightly overestimated, the position of the band edges might be right for this reaction. It is therefore not surprising that, as discussed in the Introduction, there is already a large body of experimental work trying $BiFeO_3$-based photocatalysts for the degradation of organic pollutants [19].

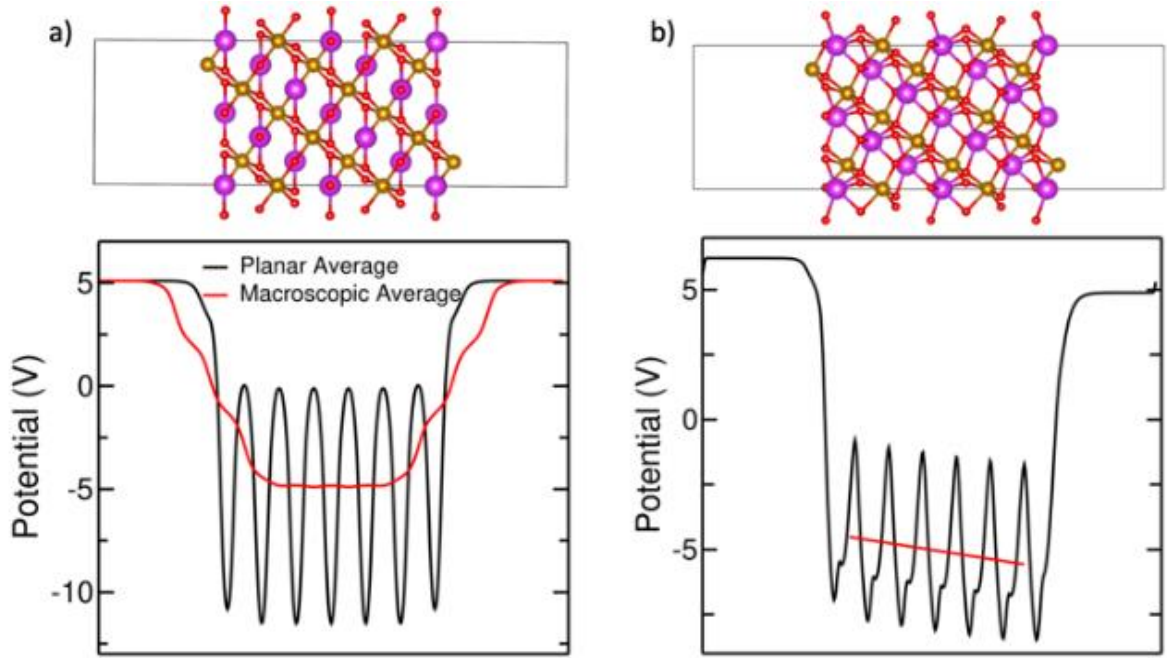

**Figure 8**: BiFeO$_3$ slabs with (111) surface terminations and the planar average electrostatic potential for (a) centrosymmetric (R-3c) phase and (b) polar (R3c) phase.

Finally, we discuss the effect of the ferroelectric polarisation on the photocatalytic properties, in a very simplified picture. The above band alignment calculation, based on a symmetric non-polar slab along the [110] direction, ignores any interplay between the band alignment and ferroelectric polarisation effects. We therefore construct another slab, now along the polarisation ([111]) direction. The surface termination of the slab is such that the only dipolar moment comes from the ferroelectric distortion: the slab with equivalent surface termination for the centrosymmetric phase (R-3c) is non-polar (**Figure 8a**). In the slab for the ferroelectric R3c phase (**Figure 8b**), the macroscopic average potential exhibits a gradient d$V$/d$z$ = 0.064 V/Å, which represents the depolarisation field that arises to compensate for the surface dipole. This depolarisation field creates a drive to separate photogenerated charge carriers, which can be expected to be beneficial for photocatalytic (and photovoltaic) applications.

The presence of the depolarisation field also has implications for the electronic band alignment, because it leads to a shift in the band positions on the surfaces of an isolated thin film, which is proportional to the magnitude of the field and to the thickness of the film. Given the calculated field intensity of ~64 mV/Å, a BiFeO$_3$ film that is, for example, ~2 nm thick in the polarisation direction, would exhibit a band shift of ~1.3 eV between its two surfaces. Such a shift is larger than the misalignment of the CB of BiFeO$_3$ for the HER. In the Co-substituted film, where the polarisation is stronger, the depolarisation field intensity will also be higher, but the shift needed to bring the CB of Co-substituted BiFeO$_3$ above the HER level is also ~0.8 eV larger than for pure BiFeO$_3$. Assuming a depolarisation field proportional to the magnitude of the polarisation, we estimate that bringing the CB of Co-substituted BiFeO$_3$ to the same level above the HER would require a thicker film, of ~4 nm.

The above discussion illustrates how polarisation may affect the photocatalytic properties of BiFeO$_3$, not only in terms of aiding charge separation but also in terms of controlling band alignment: the direction of polarisation can be used to control the position of the electronic levels at a given BiFeO$_3$ surface or interface. There are, of course, many other factors that we have not considered here. For example, in practice these films may attract free carriers from the surroundings that would partially or totally compensate the electric field. In the Co-substituted films there may be inhomogeneities of the Co distribution, potentially including phase-separated regions with higher Co concentration. The present model constitutes only an initial approximation to the theoretical modelling of what is undoubtedly a very complex phenomenon.

## 4. Conclusions

We have discussed here the response of the multiferroic material BiFeO$_3$ to Co substitution in Fe positions, considering electronic, magnetic and thermodynamic aspects, as well as potential application in photocatalysis. Co/Fe substitutions is an interesting strategy to modify the functional behaviour of BiFeO$_3$. The band gap of the system is significantly reduced as a result of Co substitution, from 2.9 eV to around 2.1 eV, and there is a simultaneous enhancing of the spontaneous polarisation. This large enhancement of ferroelectricity in this system would further promote effective carrier separation in applications such as photovoltaics or photocatalysis. Our Monte Carlo simulations show that Co$^{3+}$ ions would tend to aggregate at the concentrations studied here if an equilibrium distribution can be reached, although this phase separation is likely to be kinetically limited by cation diffusion barriers at most temperatures of interest for applications. Our calculations support the conclusion that the high electron affinity of BiFeO$_3$ makes the conduction band too negative in comparison with the level for the hydrogen evolution reaction. Therefore, unmodified BiFeO$_3$ cannot photocatalyse the full water splitting reaction. However, it can be used as a photoanode for water oxidation in photoelectrochemical cells, in combination with a suitable photocathode for the hydrogen evolution reaction. Direct Z-schemes with semiconductors with lower electron affinity could also be used for the full water splitting reaction, and we argue that Co/Fe substitutions might improve the performance of $BiFeO_3$ in such composite photocatalysts, although with some caveats in terms of band alignment and trap state effects which require further investigation. We quantify the large electric fields formed in these materials, associated with the ferroelectric distortion, and demonstrate how these fields can affect the electronic level positions. $BiCo_xFe_{1-x}O_3$ solid solutions deserve further theoretical and experimental investigation in terms of is photocatalytic applications.

**Acknowledgements**

We thank Joe Briscoe (Queen Mary University of London) and Santosh Kumar (Diamond Light Source) for useful discussion. This work made use of ARCHER, the UK's national high-performance computing service, via the UK's HPC Materials Chemistry Consortium, which is funded by EPSRC (EP/R029431), and of the Young supercomputer, via the UK's Materials and Molecular Modelling Hub, which is partially funded by EPSRC (EP/T022213/1). The collaboration between the UoR and the JNCASR was funded by an EPSRC institutional grant for international partnerships (EP/W524268/1). We acknowledge support from COST Action 18234 "Computational materials sciences for efficient water splitting with nanocrystals from abundant elements". SG is grateful for a doctoral studentship from the Felix Trust. UVW acknowledges support from a JC Bose National Fellowship of SERB-DST, Government of India.

**Conflicts of Interest**

There are no conflicts of interest to declare.

**Notes and References**

‡ Input and output files for our simulations are available in Zenodo http://dx.doi.org/10.5281/zenodo.5720628 and the associated codes are available on https://github.com/shivani-grover/Polarisation-from-BEC.